\documentclass[twocolumn,secnumarabic,amssymb,graphics,floatfix,nofootinbib,tightenlines,nobibnotes,aps,notitlepage,superscriptaddress,10pt]{revtex4-1}

\usepackage{graphicx}
\usepackage{dcolumn}
\usepackage{bm}
\usepackage{bigints}
\usepackage{hyperref}
\hypersetup{colorlinks=true}
\usepackage{xcolor}  

\usepackage{amsmath}
\usepackage{amssymb}
\usepackage{newfloat}
\usepackage{bigints}
\usepackage{upgreek}
\usepackage{bbm}
\usepackage{bm}
\DeclareFloatingEnvironment[name={Fig. S}]{suppfigure}

\begin{document}

\title{Few-mode geometric description of a driven-dissipative phase transition \\
in an open quantum system}

\author{Dmitry O. Krimer}
\email[]{dmitry.krimer@gmail.com}
\affiliation{Institute for Theoretical Physics, Vienna University of Technology (TU Wien), Wiedner Hauptstra\ss e 8-10/136, A--1040 Vienna, Austria, EU}
\author{Mikhail Pletyukhov}
\email[]{pletmikh@physik.rwth-aachen.de}
\affiliation{Institute for Theory of Statistical Physics, RWTH Aachen University, 52056 Aachen, Germany, EU}
\begin{abstract}
By example of the nonlinear Kerr-mode driven by a laser, we show that hysteresis phenomena in systems featuring a driven-dissipative phase transition (DPT) can be accurately described in terms of just two \emph{collective, dissipative} Liouvillian eigenmodes. The key quantities are just two components of a nonabelian geometric connection, even though a single parameter is driven. This powerful geometric approach considerably simplifies the description of driven-dissipative phase transitions, extending the range of computationally accessible parameter regimes, and providing a new starting point for both experimental studies and analytical insights.
\end{abstract}

\maketitle

Geometric effects manifest themselves in various fields of physics. Being very compact and elegant, a geometric formulation of a system's dynamics also provides a deeper insight by  revealing the physical redundancy in the description and paves the way to a topological classification of those properties that  remain stable under eventual external perturbations. In closed quantum systems, the most prominent example of a geometric effect is given by the Berry's phase \cite{berry84}, which is accumulated by a systems's wavefunction along with a trivial dynamical phase  during an adiabatic cyclic evolution of system's parameters. Its emergence relies on the gauge freedom in the choice of the wavefunction's phase. Physically, this is just the coordinate system choice for a transverse spin driven by a rotating magnetic field \cite{berry89}. Geometrically, it is interpreted as a holonomy effect of the parallel transport in a specific fibre bundle \cite{simon83}. The Berry connection form and the associated Chern number have become important tools for classifying ground state properties of topological insulators and superconductors \cite{altzirn,ryu,hasan2010,qi2011},  skyrmionic spin textures \cite{schultz2012,braun2012,berg2014}, topological photonic \cite{wang2009,hafezi2013,Lujoann2014} and optomechanical crystals \cite{eichen2009,safavi2014}.

In open systems, the question about a possible geometric description of system's dynamics is more subtle, since it is governed by a master equation for the system's reduced density matrix, in which the phase gauge freedom seems to disappear at first sight. Sarandy and Lidar \cite{ls1,ls2} developed a formal approach valid for slowly varying Lindblad superoperators when the time evolution of the density matrix can be represented in terms of independently evolving Jordan blocks associated with degenerated eigenvalues and treated the time-local master equation in analogy by the Schr\"odinger equation for closed systems. However, a direct transfer of closed system insights is generically obstructed by the dissipative character of open system dynamics, which is reflected in properties of the Liouvillian supermatrix. First, it is not hermitian, which implies that in general its left and right eigenvectors are not related to each other by the hermitian conjugation, even though they correspond to the same (complex-valued) eigenvalue. Second, the Liouvillian supermatrix possesses a (usually nondegenerate) zero eigenvalue, whose right eigenvector corresponds to a unique steady state, and the left one is independent of system parameters ensuring the trace conservation during the time evolution. Hence, {\it a gauge freedom in the steady state is absent}, and an observation of geometric effects, in close analogy to closed systems, is impossible in the steady state \cite{weg1}.

There are, however, other possibilities to retrieve geometric structures in open system dynamics. For example, they can emerge in the system's response functions \cite{AvronJStatPhys,albertPRX} or in observable quantities like, e.g., a pumped charge through a quantum dot \cite{weg3}; see Ref.~\cite{weg1} for a review of different approaches to the geometric description of open systems and the related effects. Moreover, dissipation can also serve as a powerful resource for generating topologically protected edge states \cite{diehl} or it can lead to nonadiabatic dynamics upon encircling of a topologically nontrivial exceptional point \cite{Rabl2015,Rotter2016}.  In this context, it is desirable to understand conditions under which geometric effects can arise in open systems. 

In this Letter, we demonstrate how the Sarandy-Lidar geometric connection arises in a more general context of open quantum systems undergoing the driven DPT \cite{carmichael} and having no exact degeneracy. As it will be demonstrated later, this connection is clearly manifested in experimentally measurable quantities. We study a basic open quantum system, the so called Kerr nonlinearity model \cite{DW}, and show that the geometric description is adequate to experimental protocols in which a characteristic time scale $t_c$ of a parameter change is comparable to the lifetime of metastable states present in this system. The importance of metastable states in the understanding of DPT has been recently realized in Refs.~\cite{plenio,Minganti:2018aa,Macieszczak:2016aa}. Here we establish the powerful approach in which the metastable dynamics can be efficiently treated in terms of just two Liouvillian modes and two components of the Sarandy-Lidar geometric connection despite the fact that the dynamics is emphatically strongly nonadiabatic relative to the smallest internal timescale generated by DPT. The gauge freedom in this context is unrelated to phase factors being provided instead by rescaling of the lengths of the Liouvillian mode vectors \cite{weg1}. The corresponding holonomy effect results in nonzero hysteresis areas in parametric dependence of observable quantities. We argue that our description is generally applicable to {\it all open quantum systems} featuring metastable dynamics near DPT, and that it sheds new light on the relation between multivalued solutions of nonlinear semiclassical (mean-field) equations of motion \cite{Armen2006,siddiqi,shumeiko,Angerer2017,Krimer:2019aa} and the linear master equation with a unique parametric solution for the density matrix. 

\begin{figure}[t!]
\includegraphics[angle=0,width=0.75\columnwidth]{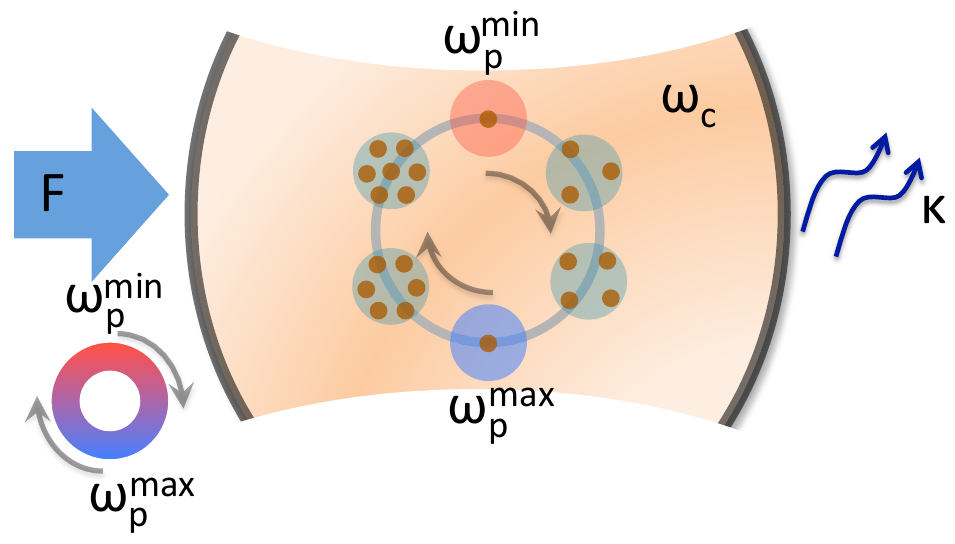}
\caption{Single-mode cavity (frequency $\omega_c$, loss rate $\kappa$) pumped by a laser field (amplitude $F$, frequency $\omega_p$). Hysteretic behavior of the cavity occupation number $n$ under a cyclic sweeping of $\delta = \omega_p - \omega_c$ is sketched by brown symbols inside circles. Red/blue colormap illustrates the sweeping range $\omega_p^{\textup{min}} \leq \omega_p \leq \omega_p^{\textup{max}}$.}
\label{fig_model}
\vspace*{-0.5cm}
\end{figure}

We exemplify our idea by considering the Kerr nonlinearity model (see Fig.~\ref{fig_model}), which consists of a weakly anharmonic ($|U| \ll \omega_c$), high-quality ($\kappa \ll \omega_c$) single-mode cavity pumped by a strong ($F \gg \kappa$) and possibly off-resonant laser field (detuning $\delta =\omega_p - \omega_c \neq 0$). Here $U$ is the strength of the Kerr self-interaction of the cavity photons, $\omega_c$ and $\omega_p$ are the cavity and pump frequencies, $\kappa$ is the cavity loss rate, and $F$ is the laser field amplitude.  The system's dynamics is governed by the Lindblad master equation
${\dot \rho} (t) = -i [ {\cal H}, \rho (t) ] +  \kappa {\cal D} [b] \rho (t) \equiv - i {\cal L} \rho (t)$ for the cavity reduced density matrix $\rho(t)$, which is expressed in the rotating frame in terms of the quantum Hamiltonian ${\cal H} = - \delta b^{\dagger} b + U/2\cdot b^{\dagger\, 2} b^2 + F (b+b^{\dagger})$ and the dissipative superoperator ${\cal D}[b] \rho(t) = b \rho b^{\dagger} - \frac12 b^{\dagger} b \rho -\frac12 \rho b^{\dagger} b$. Here $b^{\dag}$ and $b$ are the bosonic creation and annihilation operators of the cavity photons. 
 
Vectorizing $\rho (t)$, one can find the steady state solution $| \rho_{\textup{ss}} ) = \lim_{t \to \infty} | \rho (t))$ from the linear equation ${\cal L} | \rho_{\textup{ss}} )=0$, where ${\cal L}$ is the Liouvillian supermatrix and the Dirac-like bra $( \bullet |$ and ket $| \bullet )$ notations for the Liouvillian left and right eigenmode vectors are introduced. For the cases of interest, the Liouvillian supermatrix possesses a zero eigenvalue, and the steady state solution is unique and independent of initial conditions. The nonzero Liouvillian eigenvalues $\lambda_q = \omega_q - i \gamma_q$ labelled by $q=1,2, \ldots$ that result from the left $ (\bar{\rho}^q| {\cal L} = \lambda_q ( \bar{\rho}^q|$ and right ${\cal L} | \rho^q) = \lambda_q | \rho^q)$ eigenvalue problems, govern the system's dynamics $| \rho (t)) = | \rho^0) + \sum_{q \neq 0} |\rho^q ) e^{- i \lambda_q t} (\bar{\rho}^q | \rho (t=0))$ towards the steady state $| \rho_{\textup{ss}} ) \equiv | \rho^0)$. Here $\omega_q$ are the oscillation frequencies and $\gamma_q > 0$ are the relaxation rates. In the following we order eigenvalues $\lambda_q$ according to the increasing value of $\gamma_q$ (i.e., $\lambda_1$ has the smallest value of $- \textup{Im} \, \lambda_1 = \gamma_1$). Left and right eigenvectors obey the bi-orthogonality relation $(\bar{\rho}^{q'} | \rho^q) =\delta_{q' q}$, where $(\bar{\rho}^{q'} | \rho^q)$ = $\textup{tr} [(\bar{\rho}^{q'})^{\dagger} \rho^q ]$ is the Hilbert-Schmidt scalar product. The left eigenvector belonging to the zero eigenvalue is independent of the system parameters and guarantees the trace preservation $(\bar{\rho}^0| \rho (t)) = \textup{tr} [\rho (t)]$=1 (bi-orthogonality relation).

\begin{figure}[t!]
\includegraphics[angle=0,width=1.0\columnwidth]{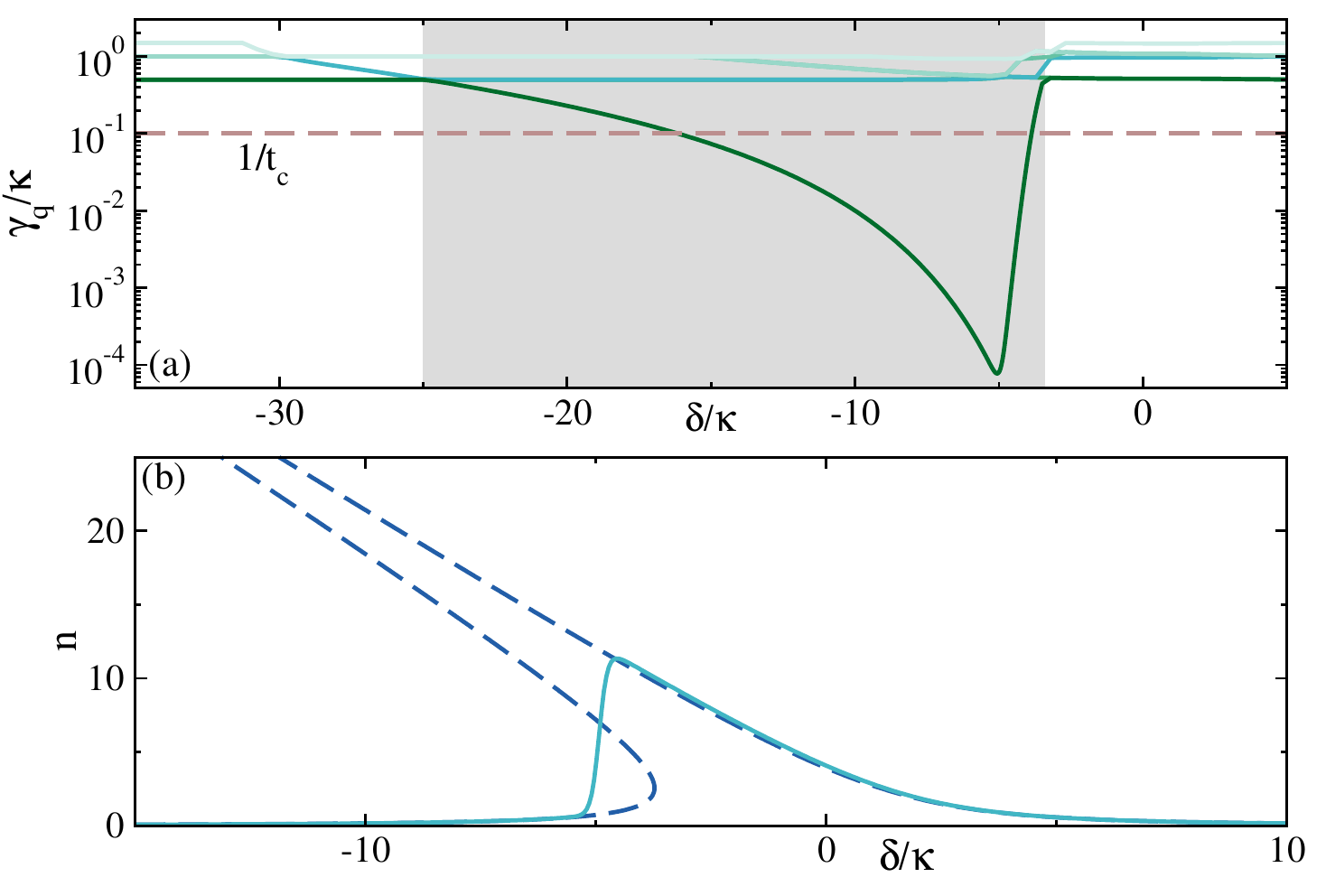}
\caption{(a) $\gamma_q$ for few values of $q$. {\it Dashed line}: $t_c=10/\kappa$ is the duration of a step used in a staircase protocol sketched in the upper panels of Fig.~\ref{fig_bdag_b_N10_50}. Gray area indicates the critical region of the width  $\approx\! 21.8\, \kappa$. 
(b) {\it Solid line}: the cavity occupation number $n=\langle b^\dag b\rangle$ for the quantum steady state versus detuning $\delta$. {\it Dashed lines}: the mean-field solution $|\langle b\rangle|^2$ shown in the same range of $\delta$. The parameter values $U=-0.5$ and $F=4$ in the units of $\kappa$.}
\vspace*{-0.4cm}
\label{fig_semicls_bdad_b_spectr}
\end{figure}

At certain parameter values the Liouvillian spectrum $\{\lambda_q | q=0,1, \ldots\}$ undergoes a nearly closing of the dissipative gap \cite{ciuti1}: $\gamma_1$ becomes vanishingly small as compared to $\kappa$, though remaining finite; see Fig.~\ref{fig_semicls_bdad_b_spectr}(a) for illustration. (Note that the gap is completely closed only in the effective thermodynamic limit featuring the DPT \cite{Casteels:2017aa,Minganti:2018aa}.) The system enters the regime of the critical slowing-down characterized by the presence of long-lived metastable states \cite{plenio}. At the same time, the steady mean-field solution, $|\beta_{ss}|^2= |\langle b  \rangle|^2$, which satisfies the nonlinear algebraic equation, $|\beta_{\textup{ss}} |^2 \left(|\beta_{\textup{ss}}|^2 - \delta/U \right)^2 +  |\beta_{\textup{ss}} |^2 \left( \kappa/2U \right)^2=F^2/U^2$, is multi-valued (the so-called optical bistability \cite{DW}). As a result, its shape considerably deviates from that of its quantum counterpart $n=\textup{tr} [ b^{\dagger} b \rho_{\textup{ss}}]$ (mostly in the critical regime of the control parameter $\delta$) which is always unique, see Fig.~\ref{fig_semicls_bdad_b_spectr}(b), but in the effective thermodynamic limit it is characterized by an infinite slope of the transition line between two different branches \cite{Casteels:2017aa,Minganti:2018aa}. In general, such a discrepancy between quantum and semiclassical solutions signals the breakdown of a mean-field description. 

%
\begin{figure}
\includegraphics[angle=0,width=0.9\columnwidth]{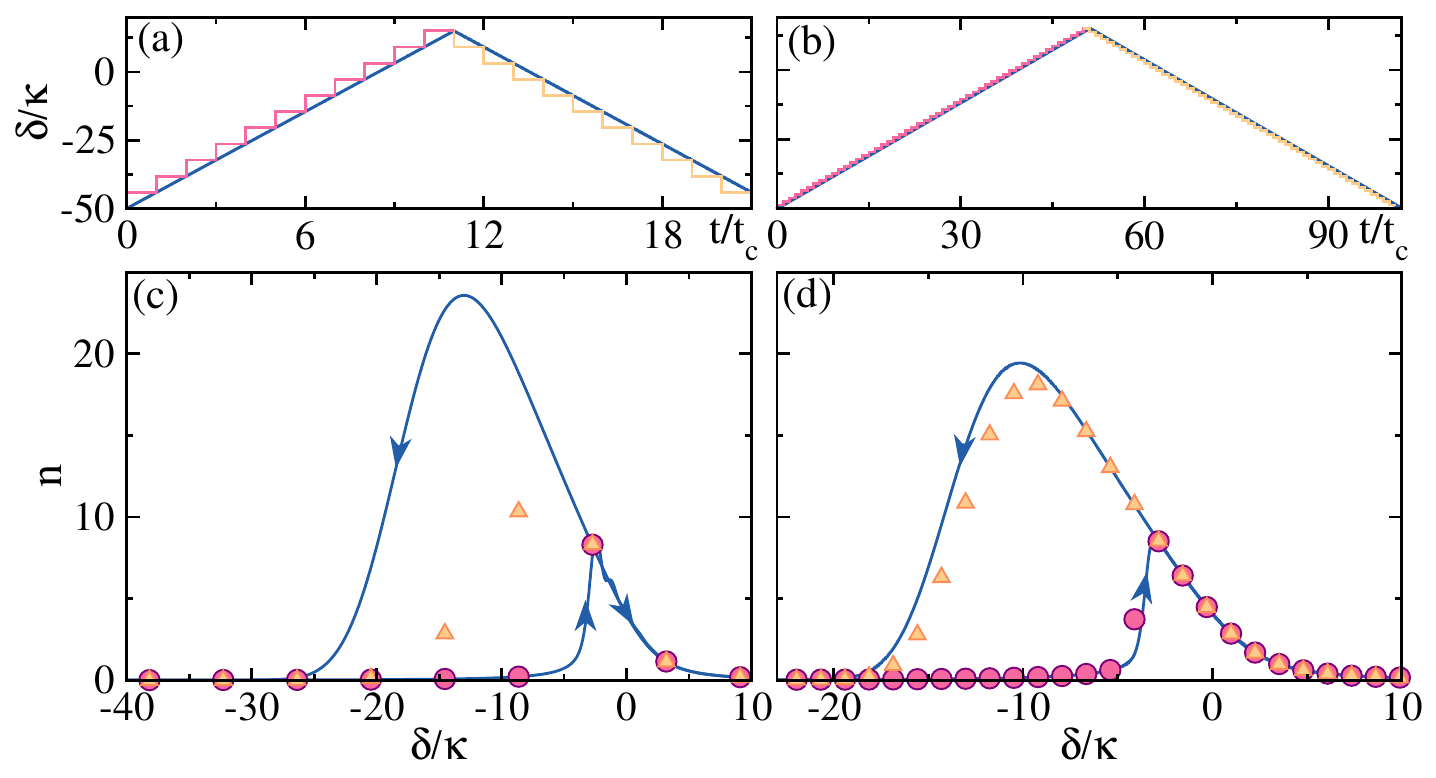}
\caption{Hysteretic behavior of the cavity occupation number $n$ {\it (lower panels)} under sweeping the detuning parameter $\delta$ through the critical region in a stepwise manner or using a linear ramp with the same inclination {\it (upper panels)}. The duration of steps used in the former protocol is always the same, $t_c=10/\kappa$ [see also the dashed line in Fig.~\ref{fig_semicls_bdad_b_spectr}(a)]. {\it Filled circles (triangles)}: values of $n$ calculated at multiples of $t_c$ at which $\delta$ abruptly changes for a forward (backward) sweep. {\it Blue lines}: corresponding numerical results for the linear ramp with arrows indicating the sweep direction. The number of steps used in the calculations for the forward and backward sweep was taken to be $N=11$ {\it (left panel)} and $N=51$ {\it (right panel)}. The values of $\kappa$, $U$ and $F$ are the same as those used in Fig.~\ref{fig_semicls_bdad_b_spectr}.}
\vspace*{-0.4cm}
\label{fig_bdag_b_N10_50}
\end{figure}

In this regard it becomes necessary to reconcile a semiclassical description of the critical multistable behaviour in a nonlinear model, which is very efficient in predicting a hysteretic effects seen in various experiments (see, e.g., \cite{devoret,Angerer2017}) and the Liouvillian description of the same system which, on one hand, captures quantum fluctuations and, on the other hand, predicts a unique steady state solution in generic models. In Refs.~\cite{ciuti1,reimer} it was proposed that the hysteresis in Markovian models can be {\it dynamically} simulated by a time-periodic change of the system's parameters with the appropriately chosen parameter ramp velocity \cite{ciuti1} or modulation frequency \cite{reimer}. Such a dynamical hysteresis has been recently observed experimentally \cite{ciuti2}. 

Here we show how a complete reconciliation of semiclassical and Liouvillian ways of describing critical phenomena in open quantum systems is naturally achieved by introducing into the theory of additional time scales characterizing a typical duration of an experiment and other scales associated with parameter ramps and measurement protocols. In particular, we introduce the time scale $t_c$, which is preset as the waiting time between two successive parameter changes. In the parameter regime close to the driven DPT, where a clear  separation $\gamma_1 \ll \gamma_{q \geq 2}$ of the relaxation rates takes place, it is meaningful to choose $t_c$ in such a way that  $\gamma_1 \ll  1/t_c \ll \gamma_{q \geq 2}$, as shown in Fig.~\ref{fig_semicls_bdad_b_spectr}(a). Choosing the quantum steady state as the initial condition just at the entrance $\delta_{\textup{min}}$ to the critical region (left edge of the gray area \cite{foot2}), we consider the first experimental run by abruptly changing the control parameter to the value $\delta_{\textup{min}} +  \Delta \delta$ and letting the system evolve during the time $t_c$ \cite{foot3}. After that the system's observables are being measured in the {\it metastable} state $| \rho^{\textup{ms}}_{\delta_{\textup{min}}+ \Delta \delta} ) \approx | \rho^0_{\delta_{\textup{min}}+ \Delta \delta}) + | \rho^1_{\delta_{\textup{min}}+ \Delta \delta}) e^{- t_c \gamma_1 (\delta_{\textup{min}}+ \Delta \delta)} ( \bar{\rho}^1_{\delta_{\textup{min}}+ \Delta \delta} |\rho^0_{\delta_{\textup{min}}})$ \cite{foot4}. Repeating this procedure $N$ times, where the number $N$ should be large enough to reach the right edge of the critical region at $\delta_{\textup{max}}$ [see Fig.~\ref{fig_semicls_bdad_b_spectr}(a)], we relate the system's metastable states at successive $(k-1)$-th and $k$-th stages of the experiment by the recursive expression   $| \rho^{\textup{ms}}_{\delta_{k}} ) \approx | \rho^0_{\delta_{k}}) + | \rho^1_{\delta_{k}}) e^{- t_c \gamma_1 (\delta_{k})} ( \bar{\rho}^1_{\delta_{k}} |\rho^{\textup{ms}}_{\delta_{k-1}})$, with $\delta_k = \delta_{\textup{min}} + k \Delta \delta$ and $1 \leq k \leq N$. Taking the observable of interest $\hat{O}$, we collect the data set of values $O_k = \textup{tr} [\hat{O} \rho_{\delta_k}^{\textup{ms}}]$, which are depicted in Fig. \ref{fig_bdag_b_N10_50}(c,d) by circles.

Analogously, we start from the value $\delta_{\max}$ and perform the same procedure with descending values $\delta'_k = \delta_{\max} -  k \Delta \delta  $. A sequence of metastable states along the reversed path in the one-dimensional parameter space is, however, different from the earlier obtained direct-path sequence even for the same values of  $\delta$. Hence the observables are also different; see again Fig. \ref{fig_bdag_b_N10_50}(c,d), where the reversed-path observable values are depicted by triangles. Although both discrete data sets are obtained with the help of the same recursive expression, the difference between them arises due to the fact that in the first case the recursion is progressive, while in the second case it is regressive. So, the hysteretic behaviour of observables in a typical experimental procedure is retrieved from the nonzero Liouvillian mode $\gamma_1$, which is responsible for the system's dynamics rather than for its steady state.


Comparing with each other the states $| \rho_{\delta}^{\textup{ms}, \pm} )= | \rho^0_{\delta}) + \chi_{\delta}^{\pm} | \rho_{\delta}^1)$, corresponding to the same value of $\delta$ and obtained from the ascending (sign "$+$") and descending (sign "$-$") staircase ramping protocols shown in Fig.  \ref{fig_bdag_b_N10_50}(a,b), we notice that the real-valued coefficients $\chi^+_{\delta}$ and $\chi^-_{\delta}$ depend on a gauge choice. The gauge freedom in the present setting is provided \cite{ls1,ls2,weg1} by a possibility to simultaneously rescale the left, $(\bar{\rho}_\delta^q |\to \frac{1}{g_{q,\delta}} (\bar{\rho}_\delta^q |$, and the right, $|\rho_\delta^q ) \to g_{q,\delta} |\rho_\delta^q)$, Liouvillian eigenvectors with $q\geq 1$ by the reciprocal real-valued coefficients. These transformations leave invariant the bi-orthogonality relations. For $q=0$ this freedom is not provided, since the rescaling of $( \bar{\rho}^0|$ is forbidden by the trace preservation; for this reason the coefficient in front of $| \rho^0_{\delta})$ in the above linear superposition is always fixed at the constant unit value, which reflects the fundamental impossibility to acquire any geometric effects as well as hysteretic phenomena in unique steady state \cite{weg1}. The gauge invariance of observables dictates the accompanying gauge transformation law $\chi_{\delta}^{\pm} \to \frac{1}{g_{1,\delta}} \chi_{\delta}^{\pm}$, which renders the ratio $\chi_{\delta}^-/ \chi_{\delta}^+$ manifestly gauge invariant. 

It is more transparent to discuss the properties of $\chi_{\delta}^\pm$ in the continuum limit. Keeping $t_c$ fixed, we reduce the discrete step $\Delta \delta$, and therefore we need to perform more steps to reach the value $\delta_{\textup{max}}$. Since the width of the critical area $(\delta_{\textup{max}} - \delta_{\textup{min}})$ is also fixed, the increment of $N$ means diminishing of a {\it ramp velocity}  $v= \Delta \delta / t_c = (\delta_{\textup{max}} - \delta_{\textup{min}})/ (N t_c)$. We notice that at sufficiently large $N \sim 50$  the difference between the staircase and the straightly linear (blue line in Fig. \ref{fig_bdag_b_N10_50}(a,b)) protocols begins to disappear: The deviation of circles and triangles collected during the staircase ramp from the continuous blue curve, which is obtained by a brute-force numerical integration of the time-dependent master equation with the linear ramp in the spirit of Ref. \cite{ciuti1}, becomes smaller and smaller with increasing value of $N$ [compare Figs.~\ref{fig_bdag_b_N10_50}(c) and \ref{fig_bdag_b_N10_50}(d)]. 

\begin{figure}[t!]
\includegraphics[angle=0,width=1.\columnwidth]{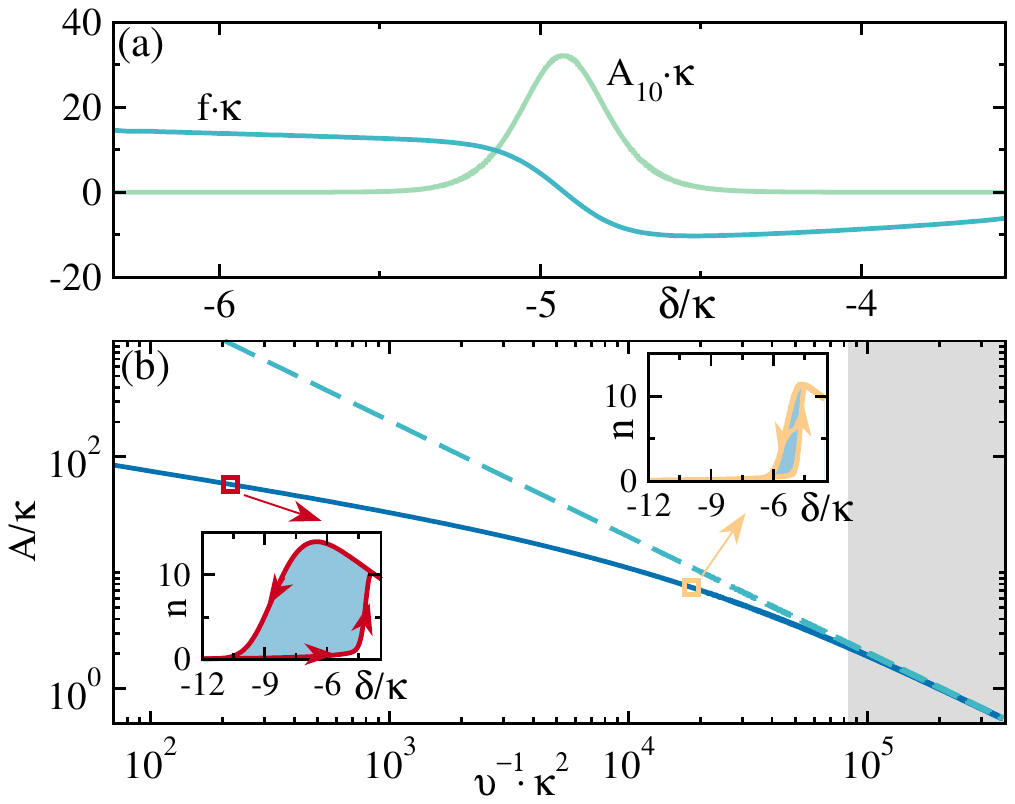}
\caption{(a) Sarandy-Lidar geometrical connection $A_{10} (\delta)$  in the gauge $\textup{tr} [b^\dag b \rho_1(\delta)]=1$ and the gauge-invariant function $f(\delta)$. (b) Hysteretic area ${\cal A}$ ({\it solid line}) as a function of $v^{-1}$. {\it Insets:} The cavity occupation number $n$ of the metastable states for the staircase protocol (see {\it upper panels} in Fig.~\ref{fig_bdag_b_N10_50}) with the number of steps $N \approx 1.4\cdot10^3$ ({\it left inset}) and $N\approx 1.2\cdot 10^5\!$ ({\it right inset}) in either direction.  {\it Dashed line}: quasiadiabatic approximation for  ${\cal A} \propto v$ becomes valid only for extremely slow sweeping velocities (gray region).}
\vspace*{-0.4cm}
\label{fig_bdag_b_MS_II}
\end{figure}

An effective geometric description of the continuum limit $\Delta \delta \to 0$ in our system featuring the driven DPT is achieved in terms of the ordinary differential equations for the gauge-invariant quantities $y^{\pm} (\delta) = \chi^{\pm} (\delta)/A_{10} (\delta)$ (see \cite{suppl} for details) 
\begin{align}
\frac{d}{d \delta} y^{\pm} (\delta) = - 1 - \left[ f(\delta) \pm \frac{\gamma_1 (\delta)}{v} \right] y^{\pm} (\delta),
\label{chi_eq}
\end{align}
expressed in terms of the {\it gauge-invariant} function $f(\delta) = A_{11} (\delta) + \frac{d}{d \delta } \ln A_{10} (\delta)$ [see Fig.~\ref{fig_bdag_b_MS_II}(a)] and equipped with initial conditions $y^+ (\delta_{\textup{min}}) = y^- (\delta_{\textup{max}}) =0$. Note that all details about the driving protocol are now hidden in a value of the ramp velocity $v$ which acquires the meaning of a global characteristic in this limit being defined as $v= (\delta_{\textup{max}} - \delta_{\textup{min}})/ (N t_c)$. This indicates that all results of Eq.~(\ref{chi_eq}) remain robust to any experimental imperfections of a driving protocol.

The Sarandy-Lidar connections $A_{q'q} (\delta ) = (\bar{\rho}^{q'} (\delta) |d/d \delta | \rho^q (\delta)) $ entering this equation are the geometric objects obtained from a solution of the instantaneous Liouvillian eigenvalue problem, in close analogy to the derivation of the Berry connection. It should be stressed that $A_{q'q} (\delta )$ can be unambiguously extracted from measurements by performing the forward and backward sweeping, see \cite{suppl} for details. The term $\propto v^{-1}$ in Eq.~\eqref{chi_eq} is the open-system analog of the dynamical phase in closed systems. The sign difference in this equation determines the hysteresis effect in the continuum limit: Experimentally, the best candidate to probe geometric effects in open systems is a hysteresis area ${\cal A} = \int_{\delta_{\textup{min}}}^{\delta_{\textup{max}}} d \delta  [\chi^- (\delta) - \chi^+ (\delta) ]$ (here expressed in the {\it observable} gauge fixed by the condition $\textup{tr} [\hat{O} \rho^1_{\delta}] =1$) for the expectation value of an observable $\hat{O}$ obtained under a cyclic variation of control parameters; see the insets in Fig. \ref{fig_bdag_b_MS_II}(b) for typical results. 

Equation~(\ref{chi_eq}) can be cast to the matrix form \cite{suppl} revealing the nonabelian character of this open system geometric effects. For this reason it is impossible to fully separate geometric and dynamical contributions to $\chi^{\pm}$. We also note that the nonabelian description provided here is substantially different from the one considered in Refs.~\cite{ls1,ls2} because in our case there is no degeneracy between $\lambda_0 =0$ and $\lambda_1 = - i \gamma_1$ Liouvillian eigenvalues. The absence of degeneracy prevents a cancelation of contributions to the hysteresis area accumulated in the forward and backward integration directions. This essential property together with the relaxation rate's separation near the DPT allows us to introduce the effective geometric description of the open system metastable dynamics even in the {\it one-dimensional} parameter space - the property, which is absent in a geometric description of the closed adiabatic dynamics.

Yet another advantage of the description in terms of Eq.~\eqref{chi_eq} is a numerically cheap study of slow and ultraslow ramps, for which a brute-force numerical study becomes very expensive. One just needs to store the values of $A_{10} (\delta)$ and $A_{11}(\delta)$ in some (e.g., observable) gauge shown in Fig. \ref{fig_bdag_b_MS_II}(a) (with $\hat{O}= b^{\dagger} b$) along with the gauge-independent function $\gamma_1 (\delta)$ shown in Fig. \ref{fig_semicls_bdad_b_spectr}(a), and solve the ODEs \eqref{chi_eq} with arbitrary small values of $v$. Thereby it is easy to obtain the scaling behavior of the hysteresis areas covering a range of sweep velocities $v$ within many orders of magnitude [see Fig.~\ref{fig_bdag_b_MS_II}(b)]. Remarkably, the quantities defining $\chi^{\pm} (\delta)$ can be extracted from a realistic experiment. To this end one needs to perform measurements at three different values of  $v$ in forward and backward direction: This suffices to find $\gamma_1 (\delta)$, $A_{10} (\delta)$, and $f(\delta)$ inside the critical region \cite{suppl}.

At ultraslow $v$ [for the parameters of Fig.~\ref{fig_bdag_b_MS_II}(b) this corresponds to $N \sim 10^6$] the dynamical term in Eq.~\eqref{chi_eq} dominates, allowing us to find the asymptotic formula $\chi^{\pm} (\delta) = \mp  v A_{10} (\delta)/\gamma_1 (\delta) + O (v^2)$. Thereby we recover the quasi-adiabatic phase of Landsberg \cite{lands} and Ning and Haken \cite{ning}, which is also known  from two-parameter pumping problems \cite{weg1,weg3}. This formula suggests that the scaling of the hysteresis areas is asymptotically linear at small $v$ (cf. \cite{ciuti1}; see also \cite{foot5}). We also find the universal relation $\chi^- (\delta) / \chi^{+} (\delta) =-1 $ in this regime.

In summary, we provided the geometric description of the metastable dynamics of the open quantum system close to DPT and proposed the way to determine the components of the Berry-like (Sarandy-Lidar) connections necessary for this purpose in a real experiment. Experimentally, our general approach can be implemented in modern hybrid quantum systems \cite{Xiang2013,Kurizki2015} based on various physical realizations, such as qubits, mesoscopic spin ensembles, and quantum metamaterials disposed on a single chip or in Rydberg systems \cite{Letscher:2017aa}.

\section{Acknowledgment} We are grateful to Maarten Wegewijs for enlightening discussions. M.P. acknowledges financial support by the Deutsche Forschungsgemeinschaft via RTG 1995. Some of the computational results presented here have been achieved using the Vienna Scientific Cluster (VSC).

%

%


%
\newpage
\begin{widetext}
\vspace{15cm}
\setcounter{equation}{0}
\renewcommand{\theequation}{S\arabic{equation}}  
\section*{Supplementary Note 1. Master equation}

In the main article we consider a Kerr nonlinearity model with the Hamiltonian
\setlength{\abovedisplayskip}{3pt}
\setlength{\belowdisplayskip}{3pt} 
\begin{eqnarray}
H^{\textup{eff}} = - \delta b^{\dagger} b + \frac{U}{2} b^{\dagger\, 2} b^2 + F (b+b^{\dagger}),
\end{eqnarray}
where $b^{\dag}$ and $b$ are standard creation and annihilation bosonic operators of the cavity photons, $U$ stands for the self-interaction strength, $\delta=\omega_p-\omega_c$ is the detuning parameter, and $F$ is the amplitude of the laser field. The Lindblad master equation for the cavity density matrix $\rho(t)$ reads
\begin{eqnarray}
\frac{d}{d t} \rho (t) = -i [ H^{\textup{eff}}, \rho (t) ] +  \kappa D [b] \rho (t), 
\end{eqnarray}
where $D[b] \rho$ is the Lindblad dissipator given by $D[b] \rho = b \rho b^{\dagger} - \frac12 b^{\dagger} b \rho -\frac12 \rho b^{\dagger} b$. Representing $\rho (t)$ in the Fock basis, $|n \rangle$, as $\rho (t) = \sum_{m,n} \rho_{mn} (t)  |m \rangle \langle n |$, we obtain equations for the matrix elements ($m,n=0,1,2,\ldots$)
\begin{eqnarray}
\label{Eq_rho_mn}
&&- i \frac{d}{d t} \rho_{m,n} =\left[ (m-n) \delta - U (m-n) \frac{m+n-1}{2}   + \frac{i (m+n)}{2} \kappa \right]  \rho_{m,n} 
\nonumber \\
&&- i \kappa \rho_{m+1,n+1} \sqrt{(m+1) (n+1)}  +  F \sqrt{n+1} \rho_{m,n+1} \\\nonumber
&&- F \sqrt{m+1} \rho_{m+1,n} -  F \sqrt{m} \rho_{m-1,n} + F \sqrt{n} \rho_{m,n-1}.
\end{eqnarray}
To facilitate numerical calculations we represent the above equation in the following form
\begin{eqnarray}
i \frac{d }{d t } | \rho (t)) = {\cal L} | \rho(t)),
\label{liouv}
\end{eqnarray}
and introduce a cutoff $d$ in the Fock space. Then,  the $d \times d$ density matrix is converted into a column vector with $d^2$ elements by arranging all matrix elements of $\rho_{m,n}$ in accordance with the following rule
\begin{eqnarray}
\label{rho_vec}
|\rho(t)) = \rho_p \rightarrow (\rho_{11},\, \rho_{12},\,  ... ,\,  \rho_{1d},\,\, \rho_{21},\, ..., \, \rho_{2d}, \,..., \,\rho_{d1},\,...,\,\rho_{dd})^T,
\end{eqnarray}
where $p=(m-1)\times d+n$. In turn, the Liouvillian superoperator ${\cal L}$ is  represented by a $d^2 \times d^2$ matrix. Note that in Eqs.~(\ref{Eq_rho_mn},\ref{liouv}) we have explicitly introduced the imaginary $i$, which also affects a definition of the Liouvillian; this is done to obtain a formal analogy with the Schr\"odinger equation. 

We solve then Eq.~(\ref{liouv}) numerically using the standard Runge-Kutta method of the 4th order by taking a quantum stationary state as an initial condition at $t=0$, $| \rho(0))=|\rho^0)$. Next, we calculate the cavity occupation number $n(t)$ as a function of time by the formula
\setlength{\abovedisplayskip}{2pt}
\setlength{\belowdisplayskip}{2pt}
\begin{eqnarray}
n (t)=\textup{tr} [b^\dag b\rho(t)]=\sum_{k=0}^d k\, \rho_{k,k}(t),
\label{n_occup}
\end{eqnarray}
where the diagonal elements  $\rho_{k,k}(t)$ are unambiguously extracted from the vector representation (\ref{rho_vec}). Typical results of calculations are presented in Fig.~S\ref{fig_bdag_b_UP_DOWN_rough_vs_t_N11} (red curve) when sweeping the detuning parameter $\delta$ across the critical region designated by gray area in Fig.~2(a) of the main article.  (All other parameters are kept the same.) In this example we take a staircase protocol shown in Fig.~3(a) of the main article. As an initial condition at $t=0$, a quantum stationary state, $|\rho^0)$, is chosen for the detuning parameter $\delta_0$ which lies to the left with respect to the critical region. After the whole cycle the detuning parameter returns to its initial value $\delta_0$ (the final time integration in Fig.~S\ref{fig_bdag_b_UP_DOWN_rough_vs_t_N11}). One can see that all oscillations fade out on the time scale $t_c$ (at which the current value of $\delta$ is kept constant), before a subsequent jump to a new value, $\delta+\Delta \delta$, will occur. This result is in accordance with our main assumption that the contributions of all but the smallest nonzero eigenvalue are negligibly small for a choice of $t_c$ such that $\min(\gamma_1)<1/t_c<\kappa/2$, as shown in Fig.~2(a) of the main article (see also orange symbols in Fig.~S\ref{fig_bdag_b_UP_DOWN_rough_vs_t_N11} and discussions in the next section).

\section*{Supplementary Note 2. Eigenvalue problem and metastable states}

Here we briefly present details of solving the left and right eigenvalue problems for Liouvillian supermatrix and introduce the concept of metastable states. 

In  space of density matrices, the Hilbert-Schmidt  scalar product is defined by $(A | B) = \textup{tr} [A^{\dagger} B]=(A | B) = \sum_{a,b=1}^d A^*_{ab} B_{ab}$. Introducing a multi-index $q=(ab)$ (which also reflects a way of re-arranging a matrix $A$ into a column vector $|A)$), we obtain a usual form of the scalar product $(A | B) = \sum_{q=0}^{d^2-1} A^*_{q} B_{q}$. 

The right eigenvalue problem associated with \eqref{liouv} at a certain fixed value of some control parameter $r$ reads (for the sake of simplicity a parametric dependence on $r$ is omitted below)
\begin{eqnarray}
 {\cal L} | \rho^q ) = \lambda_q | \rho^q),\,\,\,  \textup{or} \quad \sum_{s=0}^{d^2-1} {\cal L}_{ps} \rho^q_s = \lambda_q \rho^q_p,
\label{rightEVP}
\end{eqnarray}
where $q=0, \ldots , d^2 - 1$. Here $\lambda_q = \omega_q - i \gamma_q$ are $d^2$ (eventually nondegenerate) complex-valued eigenvalues enumerated in the ascending order of their imaginary parts. (For the model of our interest, a validity of the nondegeneracy assumption is checked numerically.) Note that a physically meaningful ${\cal L}$ must have zero eigenvalue, $\lambda_0 =0$, and the corresponding right eigenvector $| \rho^0)$, realizes the stationary state of a system in the long-time limit. For all other eigenvalues, the causality condition $\gamma_q > 0$ must hold (see Fig.~2(a) of the main article). 

In contrast to closed systems, right and left eigenvalue problems for the dissipative systems have to be solved separately, as left eigenvectors cannot be obtained from right eigenvectors by means of hermitian conjugation. Therefore, we also numerically solve the left eigenvalue problem
\begin{eqnarray}
 (\bar{\rho}^q |  {\cal L}   = \lambda_q ( \bar{\rho}^q |,\,\,\, \textup{or} \quad \sum_{p=0}^{d^2-1} (\bar{\rho}^q_p)^*  {\cal L} _{ps} = \lambda_q(\bar{\rho}^q_s)^* .
\label{leftEVP}
\end{eqnarray}
In practice we express Eq.~\eqref{leftEVP} in a more convenient form of an effective right eigenvalue problem $\sum_{l=0}^{d^2-1} ({\cal L}^T)_{sp}  (\bar{\rho}^q_p)^* = \lambda_q(\bar{\rho}^q_s)^*$. From this representation it becomes clear why we have the same eigenvalues in the right \eqref{rightEVP} and the left \eqref{leftEVP} eigenvalue problems: ${\cal L}$ and ${\cal L}^T$ have the same characteristic equations, and therefore the same spectra. From \eqref{rightEVP} and \eqref{leftEVP} we obtain
\begin{eqnarray}
 \sum_{p=0}^{d^2-1}  \sum_{s=0}^{d^2-1} (\bar{\rho}^q_p)^*  {\cal L} _{ps} \rho^{q'}_s= \lambda_q  \sum_{m=0}^{d^2-1}  (\bar{\rho}^q_s)^*  \rho^{q'}_s ,
\end{eqnarray}
which yields $\lambda_{q'} (\bar{\rho}^q | \rho^{q'}) =  \lambda_q (\bar{\rho}^q | \rho^{q'})$. This implies that for $q \neq q'$ (and hence for $\lambda_q \neq \lambda_{q'}$ because of the nondegeneracy) it holds $ (\bar{\rho}^q | \rho^{q'})=0$. In turn, for $q=q'$, one can ever normalize the state $(\bar{\rho}^q |$ to enforce the bi-orthogonality condition 
\begin{eqnarray}
(\bar{\rho}^q | \rho^{q'})  = \delta_{qq'}.
\label{biorth}
\end{eqnarray}
Then, we can introduce a spectral decomposition of the Liouvillian
\begin{eqnarray}
{\cal L} = \sum_{q=0}^{d^2-1} \lambda_q | \rho^q) ( \bar{\rho}^q | ,\,\,\,  \textup{or} \quad {\cal L}_{ps}  = \sum_{q=0}^{d^2-1} \lambda_q \rho^q_p (\bar{\rho}^q_s)^* ,
\label{resol_L}
\end{eqnarray}
as well as the identity resolution
\begin{eqnarray}
 \mathbbm{1}_{d^2 \times d^2} = \sum_{q=0}^{d^2-1}  | \rho^q) ( \bar{\rho}^q | , \,\,\, \textup{or} \quad \delta_{ps} =  \sum_{q=0}^{d^2-1} \rho^q_p(\bar{\rho}^q_s)^*.
\end{eqnarray}
It should be emphasized that the trace conservation of the density matrix corresponding to the stationary state acquires the following form in these notations 
\begin{eqnarray}
(\bar{\rho}^0 | \rho^{0}) = \sum_{l=0}^{d^2 -1} ( \bar{\rho}^0_s)^* \rho^{0}_s =1.
\label{left0}
\end{eqnarray}
Therefore, the eigenvector $(\bar{\rho}^0 |$ has components $ \bar{\rho}^0_s =1$ for all "diagonal" $s=(aa)$, and $ \bar{\rho}^0_s =0$ for all "nondiagonal" $s = (ab)$, $a \neq b$. Thus, $(\bar{\rho}^0 |$ has no any parametric dependence, and its action on the right vector $|\rho(t))$ is  equivalent to applying the trace operation, i.e. $(\bar{\rho}^0 | \rho(t) )=\textup{tr}[\rho(t)]=1$. Note also that in Eq.~\eqref{resol_L} the summation over $q$ effectively starts from $q=1$, since $\lambda_0=0$.

It is worth noting that one has the gauge freedom in defining right and left Liouvillian eigenmodes belonging to the same eigenvalue $\lambda_q \neq 0$ which is expressed in their simultaneous rescaling by reciprocal scaling factors,
\begin{eqnarray}
| \rho^{q} (r)) \to | \rho^{q} (r)) g_q (r), \quad  (\bar{\rho}^q (r) |  \to \frac{1}{g_q (r)} (\bar{\rho}^q (r)|,
\label{Eq_gauge_freed}
\end{eqnarray}
preserving the bi-orthogonality relations (\ref{biorth}). The only exception is $q=0$ with $\lambda_0=0$: In this case the corresponding right eigenmode corresponds to unique stationary state, while the left one is nothing else but the trace taking operation. Since the trace of a physical density matrices must be always  preserved, it follows that the rescaling gauge freedom is absent for zero eigenmodes of the Liouvillian.

Importantly, the Liouvillian spectral decomposition \eqref{resol_L} enables us to rewrite the formal solution of Eq.~\eqref{liouv}, $ | \rho (t)) =  e^{- i {\cal L} t} | \rho (0))$, in a form which is very useful  for a direct implementation of numerical calculations
\begin{eqnarray}
| \rho (t) ) = |\rho^{0}) +  \sum_{q=1}^{d^2-1}   | \rho^q) ( \bar{\rho}^q  | \rho (0) ) e^{-i \omega_q t - \gamma_q t}.
\label{rho_evol}
\end{eqnarray}

Suppose now that some control parameter $r$ changes in a stepwise manner such that it jumps by the same amount of $\Delta r$ after equal time steps $t_c$ (in the main article $r\equiv\delta$, see upper panels of Fig.~3 therein). We can easily adapt Eq.~(\ref{rho_evol}) for this protocol deriving the following iterative formula which connects two subsequent parameter jumps, $r_0+(k-1)\Delta r$ and $r_0+k\Delta r$ as follows
\begin{eqnarray}
| \rho_{r_0+k\Delta r} ) = |\rho^{0}_{r_0+k\Delta r}) +  \sum_{q=1}^{d^2-1}   | \rho^q_{r_0+k\Delta r}) ( \bar{\rho}^q_{r_0+k\Delta r}  | \rho_{r_0+(k-1)\Delta r}  ) e^{-i \omega_q t_c - \gamma_q t_c},
\label{rho_evol_tc}
\end{eqnarray}
where $\omega_q$ and $-\gamma_q$ are real and imaginary parts of the eigenvalue $\lambda_q$ computed at $r_0+k\Delta r$. Note that in Eq.~(\ref{rho_evol_tc}) the eigenvalue problem as well as the scalar product $( \bar{\rho}^q_{r_0+k\Delta r}  | \rho_{r_0+(k-1)\Delta r}) $ are computed at each step of this iterative procedure. Throughout the paper, we choose a stationary state, 
$|\rho^q_{r_0})=|\rho^0_{r_0})$, as the initial condition at $r=r_0$.

\begin{suppfigure}
\includegraphics[angle=0,width=0.5\columnwidth]{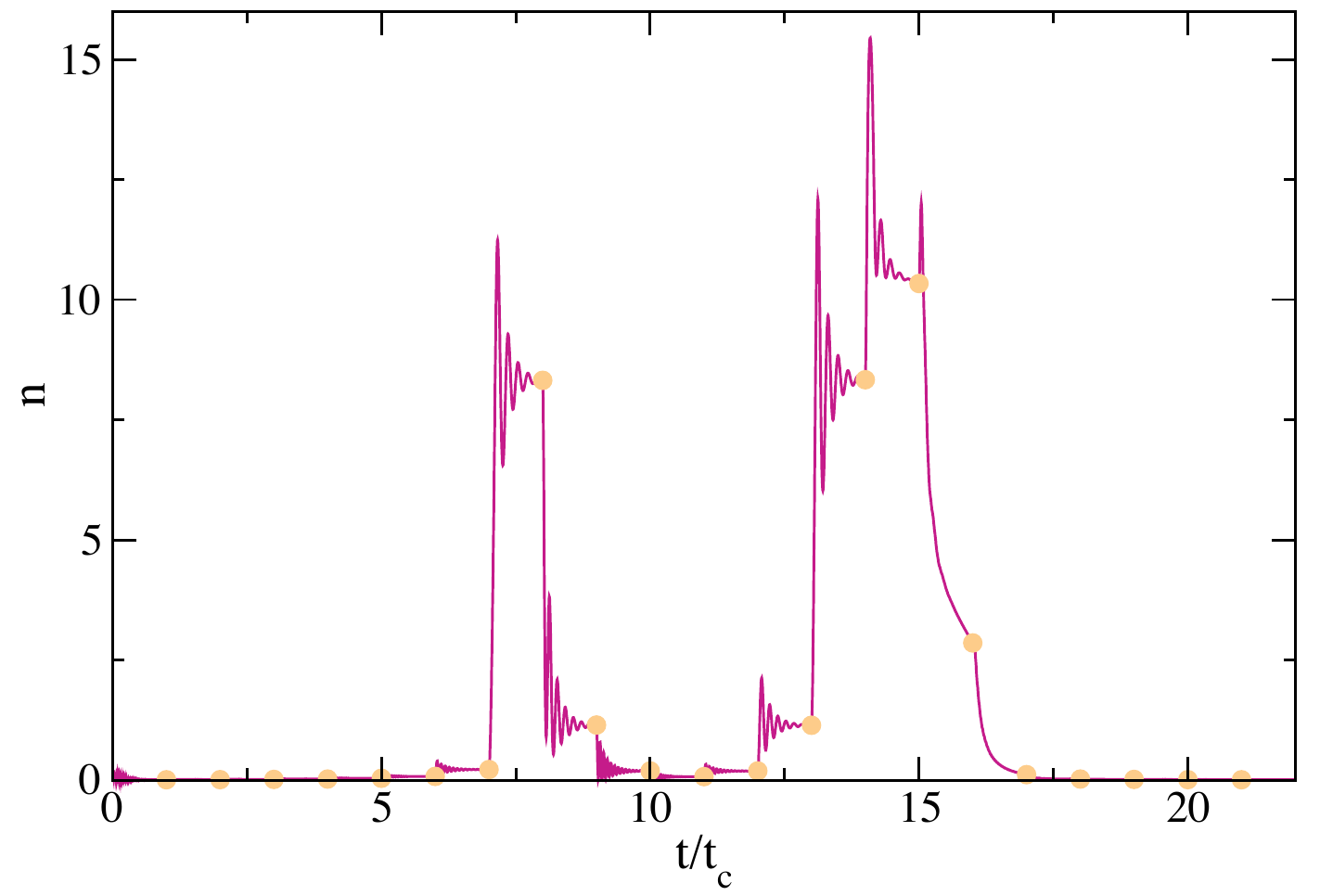}
\caption{The cavity occupation number $n$ as a function of time (in units of $t_c$) for the whole sweep cycle shown in Fig.~3(a,c) of the main article (the number of steps $N=11$ for the forward and backward sweep). {\it Red curve}: numerical solution of the time-dependent master equation. {\it Orange symbols}: values of $n$ for the metastable states given by Eq.~(\ref{rho_evol_tc_MS}) calculated at multiples of $t_c$ at which $\delta$ abruptly changes by $\pm \Delta \delta$. The cavity decay rate, the strength of nonlinearity and the driving amplitude are, respectively, $\kappa=1$, $U=-0.5$ and $F=4$.}
\label{fig_bdag_b_UP_DOWN_rough_vs_t_N11}
\end{suppfigure}

In the main paper the detuning $\delta$ plays a role of the control parameter, $r\equiv\delta$, and the time step $t_c$ is chosen such that the following inequality is satisfied, 
\begin{eqnarray}
\min(\gamma_1)<1/t_c<\kappa/2,
\label{Ineq_MS}
\end{eqnarray}
where the decay rate $\gamma_1$ is the imaginary part of $-\lambda_1$ (see Fig.~2(a) of the main article). As a consequence, we expect that the contributions from all other Liouvillian modes to the dynamics fade out during the time $t_c$, before $\delta$ jumps to a subsequent value $\delta+\Delta \delta$. (Recalling that eigenvalues are enumerated in the ascending order of absolute values of their imaginary parts.) In this case the description of the resulting dynamics is drastically simplified,  Eq.~(\ref{rho_evol_tc}) being reduced to
\setlength{\abovedisplayskip}{3pt}
\setlength{\belowdisplayskip}{3pt}
\begin{eqnarray}
| \rho_{r_0+k\Delta r}^{\textup{ms}}) = |\rho^{0}_{r_0+k\Delta r}) +  | \rho^1_{r_0+k\Delta r}) ( \bar{\rho}^1_{r_0+k\Delta r}  | \rho^{\textup{ms}}_{r_0+(k-1)\Delta r}  ) e^{-\gamma_1 t_c},
\label{rho_evol_tc_MS}
\end{eqnarray}
where the superscript "$\textup{ms}$" underscores hat the resulting state is metastable and  consists of just two Liouvillian modes, namely the zero mode and the softest one with the smallest relaxation rate  $\gamma_1$. Another important feature is that the real part of the first eigenvalue vanishes in the whole critical region, $\omega_1=0$ (gray area in Fig.~2(a) of the main article). 

To justify our expectation we calculate the cavity occupation number $n(t)$ as a function of time given by Eq.~(\ref{n_occup}) taking $| \rho_{r_0+k\Delta r}^{\textup{ms}})$ from Eq.~(\ref{rho_evol_tc_MS}) as the density matrix. The results of numerical calculations turn out to be in a perfect agreement with those obtained by integrating the master equation \eqref{liouv} with the time-dependent Liouvillian $\mathcal{L} (t)$ (through $r (t)$) (see Supplementary Note 1); compare orange symbols with the red curve in Fig.~S\ref{fig_bdag_b_UP_DOWN_rough_vs_t_N11}.

\section*{Supplementary Note 3. Simplified description of metastable states}

Here we briefly sketch the derivation of the ODE governing  the evolution of metastable states (\ref{rho_evol_tc_MS}) provided that increments $\Delta r$ are sufficiently small, as it is the case for all results presented in Fig.~4 of the main article.

\begin{suppfigure}[t!]
\includegraphics[angle=0,width=0.5\columnwidth]{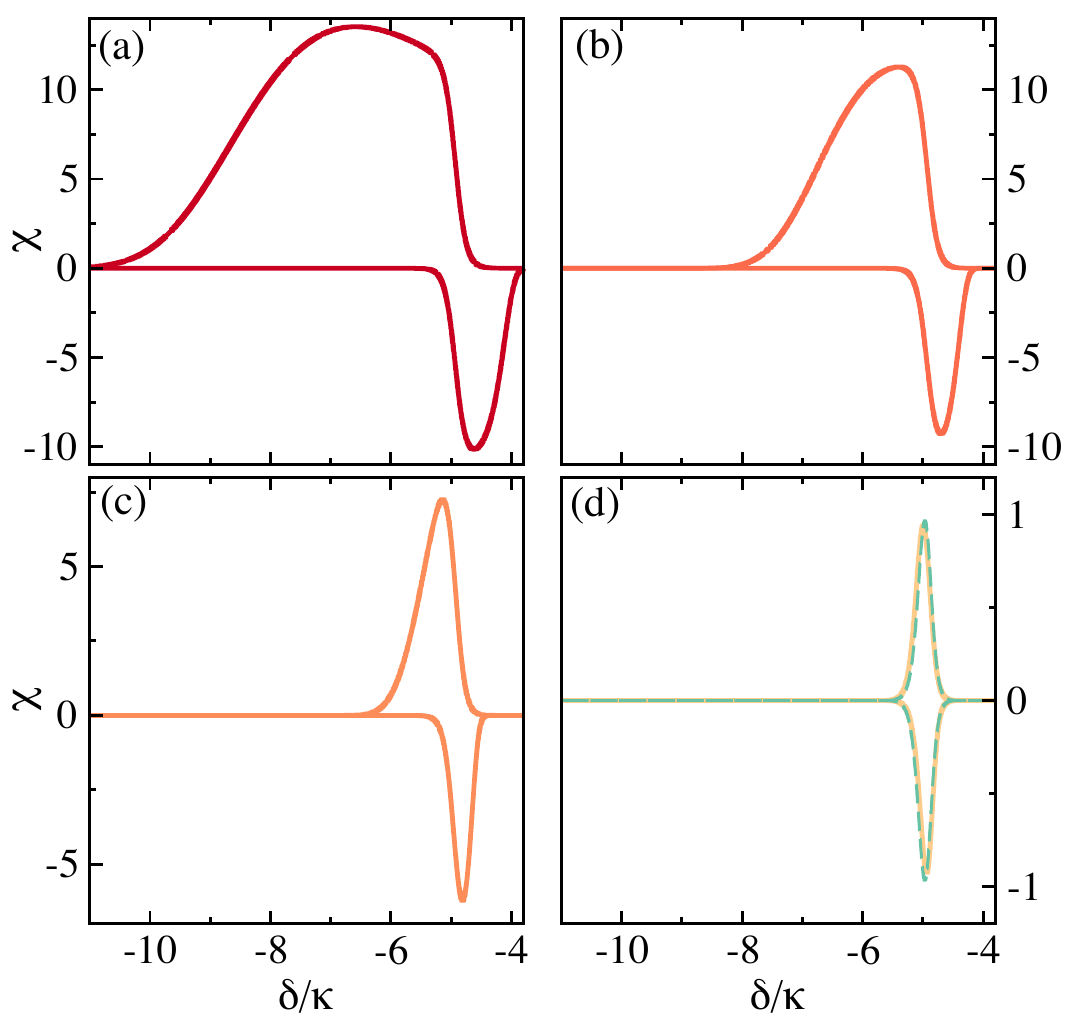}
\caption{The functions $\chi^{\pm}$, determining metastable states, versus $\delta$. They result from  Eqs.~(\ref{dchi1_dr}) solved for the values of $v = (\delta^{\textup{max}} - \delta^{\textup{min}})/(N t_c) $ corresponding to various values $N\approx 1.4\cdot10^3\!$, $1.2\cdot\!10^4\!$, $1.2\cdot\!10^5$, and $2.4\cdot10^6\!$. {\it Green dashed line}: analytical quasiadiabatic solution given by Eq.~(\ref{Eq_v_vsmall}). Metastable state values of the occupation number expressed via $\chi^{\pm}$ coincide with the numerical results of integration of the time-dependent master equation to a very good accuracy. The panels (a),(c) correspond to the insets in Fig.~4(b) of the main article.}
\label{fig_chi}
\end{suppfigure}

Expanding $| \rho (t))$ in the right instantaneous eigenbasis of ${\cal L}(t)$ as 
\begin{eqnarray}
| \rho (t)) = \sum_{p=0}^{d^2-1} \chi_p (t) | \rho^p (t))
\label{rho_Sum}
\end{eqnarray}
and substituting this expansion into Eq.~(\ref{liouv}) we obtain
\begin{eqnarray}
 \sum_{p=0}^{d^2-1} \dot{\chi}_p (t) | \rho^p (t)) +  \sum_{p=0}^{d^2-1} \chi_p  (t) \frac{d}{dt} | \rho^p (t)) = - i \sum_{p=0}^{d^2-1} \lambda_p (r(t))  \chi_p (t) | \rho^p (t)).
\end{eqnarray}
Projecting on $(\bar{\rho}^q (t)|$ and using the bi-orthogonality relation (\ref{biorth}) leads to
\begin{eqnarray}
\dot{\chi}_q (t)  +  \sum_{p=0}^{d^2-1} \chi_p  (t) (\bar{\rho}^q (t) | \frac{d}{dt} | \rho^p (t)) = - i  \lambda_q (r(t))  \chi_q (t).
\label{Eq_gen_chi}
\end{eqnarray}
Introducing the parameter change velocity $v=dr/dt$, we obtain  the following equation 
\begin{eqnarray}
\frac{d}{d r} \chi_q (r) + \sum_{p=0}^{d^2-1} A_{qp} (r) \chi_p (r) = - i \frac{1}{v} \lambda_q (r)  \chi_q (r),
\label{dchi_dr}
\end{eqnarray}
where the matrix elements $A_{qp}(r) = (\bar{\rho}^q (r) |d/dr| \rho^p (r))$ have the meaning of geometric connections, and the velocity-dependent term on the r.h.s. of Eq.~(\ref{dchi_dr}) stands for the dynamical contributions to $\chi_q (r)$.  Since $(\bar{\rho}^0|$ does not depend on $r$ (see Supplementary Note 2), it follows that $A_{0p} (r)=d/dr(\bar{\rho}^0| \rho^p (r))=0$ for each $p$ due to the bi-orthogonality condition (\ref{biorth}). Therefore, $d\chi_0 (r)/dr =0$, because $\lambda_0=0$ on the r.h.s. of Eq.~(\ref{dchi_dr}). To satisfy the initial condition at $r=r_0$, $|\rho^q_{r_0})=|\rho^0_{r_0})$, we should set $\chi_0 (r)=1$.

In the previous note we proved that if the inequality (\ref{Ineq_MS}) for time steps $t_c$ is satisfied, then contributions to the dynamics of all, but the first, eigenmodes are negligibly small. (Importantly, this rather general statement is perfectly applicable even for the case where a continuous description is not possible, see e.g. Fig.~3(a) in the main paper). For the less general case of continuous parameter change protocols, the higher Liouvillian modes  contribution to the dynamics is also negligible, and the sum in Eq.~(\ref{rho_Sum}) can be truncated
\begin{eqnarray}
|\rho (r)) \approx  | \rho^0(r))+\chi_1 (r) | \rho^1 (r))
\label{rho_r_1}
\end{eqnarray}
(recall that $\chi_0 (r)=1$). In addition, in Eq.~(\ref{dchi_dr}) for $\chi_1(r)$ all terms $A_{1p}(r)$ with $p \geq 2$ give negligibly small contribution. Therefore, Eq.~(\ref{dchi_dr}) reduces to the the following matrix form revealing the nonabelian structure of metastable dynamics 
\begin{eqnarray}
\frac{d}{d r} \left( \begin{array}{c} 1 \\ \chi^{\pm}(r)  \end{array} \right) =  -  \left( \begin{array}{cc} A_{00} \pm  \gamma_0 /v   & A_{01}(r) \\ A_{10}(r) & A_{11}(r) \pm  \gamma_1(r) /v   \end{array} \right)  \left( \begin{array}{c} 1 \\ \chi^{\pm}(r)  \end{array} \right).
\label{Eq_Matrix_form}
\end{eqnarray}
where the mode index in $\chi$ is omitted for brevity and the additional superscript $\pm$ is introduced for clarity. The required components of $A_{q'q}$ have the gauge transformation properties $A_{10}(r) \to \frac{1}{g_{1} (r)} A_{10}(r) $, $A_{11}(r) \to A_{11}(r) + \frac{d}{d r} \ln g_{1} (r)$. As mentioned above, the components $A_{0q} (r) \equiv 0$ owing to the bi-orthogonality relations and the parametric independence of  $(\bar{\rho}^0|$. In addition, $\gamma_0=0$. The signs "$\pm$" designate the ascending (sign "$+$") and descending (sign "$-$") staircase ramps shown in Fig.~3(a,b) of the main article.
The first line in Eq.~(\ref{Eq_Matrix_form}) is trivially satisfied, whereas the second line leads to the following ODEs for $\chi^{\pm} (r)$
\begin{eqnarray}
\frac{d}{d r} \chi^{\pm} (r) +  A_{10} (r) + A_{11} (r) \chi^{\pm} (r) =\mp\frac{1}{v} \gamma_1 (r)  \chi^{\pm} (r),
\label{dchi1_dr}
\end{eqnarray}
with the initial conditions $\chi^+ (r_0)= \chi^- (r_1) = 0$. Finally, we rewrite the above equation in terms of the gauge-invariant quantities $y^{\pm} (r) = \chi^{\pm} (r)/A_{10} (r)$ and $f (r) = A_{11} (r) + \frac{d}{d r} \ln A_{10} (r) $ (cf. Eq.~(1) of the main article with $r\equiv \delta$) as follows
\begin{eqnarray}
\frac{d}{d r} y^{\pm} (r) = - 1 - \left[ f(r) \pm \frac{\gamma_1 (r)}{v} \right] y^{\pm} (r).
\label{chi_eq_suppl}
\end{eqnarray}
The solutions of Eqs.~(\ref{chi_eq_suppl}) read
\begin{align}
&  y^{+} (r) =  \int_{r_{\textup{min}}}^{r} d r' e^{-\int_{r'}^{r} d r'' f (r'')} e^{-\frac{1}{v} \int_{r'}^{r} d r'' \gamma_1 (r'')}, \\
&  y^{-} (r) =  -  \int_{r}^{r_{\textup{max}}} d r' e^{\int^{r'}_{r} d r'' f (r'')} e^{- \frac{1}{v}\int^{r'}_{r} d r'' \gamma_1 (r'')}.
\end{align}
Apparently, $\chi^- (r)/ \chi^+ (r) = y^{-} (r) / y^{+} (r)$ is a gauge-invariant quantity.

For decreasing values of $v$ the dependence of $y^{\pm}$ on $f$ weakens since the dynamical term starts to dominate. Therefore, these factors become mostly dynamical, and we have an approximate factorization of $\chi^{\pm}$ into a product of the geometric part ($A_{10}$) and the dynamical part ($y^{\pm}$).

The connection components $A_{1p}(r)$ ($p=0,1$) in Eq.~(\ref{dchi1_dr}) can be found numerically by solving the eigenvalue problem (\ref{rightEVP}) at different values of $r=r_0+l \Delta r$ within the critical region (gray area in Fig.~2(a) of the main article) and by using the central difference approximation to the first derivative, namely
\begin{eqnarray}
A_{1p}(r) = (\bar{\rho}^1 (r) |\dfrac{d}{dr}| \rho^p (r)) \approx \dfrac{1}{2\Delta r}\left\{  ( \bar{\rho}^1_{r_0+l\Delta r}|\rho^{p}_{r_0+(l+1) \Delta r})-( \bar{\rho}^1_{r_0+l\Delta r}|\rho^{p}_{r_0+(l-1) \Delta r})\right\}.
\label{A_1p_fin_expr}
\end{eqnarray}
In practice, we compute $A_{1p}(r)$ at $\sim 200$ points of $r$ and then use a spline interpolation procedure to obtain the $r$-dependence within the critical region (see Fig.~4(a) in the main article). Feeding it to Eq.~(\ref{dchi1_dr}), we solve this ODE at arbitrary values of the sweep velocity $v$. 

Finally, we calculate the cavity occupation number $n^{\pm} (r)=\textup{tr} [b^\dag b\rho^{\pm} (r)]$ with help of $|\rho^{\pm} (r))$ defined by Eq.~(\ref{rho_r_1}). The resulting expression is given by $n^{\pm} (r)=n^{\textup{st}}(r)+\chi^{\pm} (r)\,  \textup{tr} [b^\dag b\rho^1(r)]$, where  $n^{\textup{st}}(r)=\textup{tr} [b^\dag b\rho^0(r)]$ is the stationary state occupation number. Exploiting the gauge freedom (\ref{Eq_gauge_freed}), we choose $g_1 (r)$ in such a way that the condition $\textup{tr} [b^\dag b\rho^1(r)]=1$ is enforced (the so called {\it observable} gauge, in this case $\hat{O} = b^{\dagger} b$). In this gauge, we obtain a simple expression
\begin{eqnarray}
n(r)=n^{\textup{st}}(r)+\chi^{\pm} (r),
\label{n_occup_chi}
\end{eqnarray}
and the functions $\chi^{\pm} (r)$ are shown in Fig.~S\ref{fig_chi}.

At small $v$, we can expand $\chi^{\pm} (r)$ in series $\chi^{\pm} (r)= \sum_{n=1}^{\infty} \chi{^{(n)\pm}} (r)\, v^n$. Inserting it into Eq.~(\ref{dchi1_dr}), we find the leading contribution 
\begin{eqnarray}
\chi^{(1) \pm} = \mp A_{10} (r)/ \gamma_1 (r),  \label{Eq_v_vsmall}
\end{eqnarray}
and hence 
\begin{align}
| \rho (r)) &= | \rho^0 (r)) \mp \frac{v}{\gamma_1(r)} A_{10}(r) | \rho^1 (r)) + O (v^2), \\
n(r) & \approx  n^{\textup{st}}(r) \mp \frac{v}{\gamma_1(r)} A_{10}(r).
\end{align}
Graphically, the results in this quasiadiabatic regime are presented in Fig.~S\ref{fig_chi}(d).

\begin{suppfigure}[t!]
\includegraphics[angle=0,width=0.75\columnwidth]{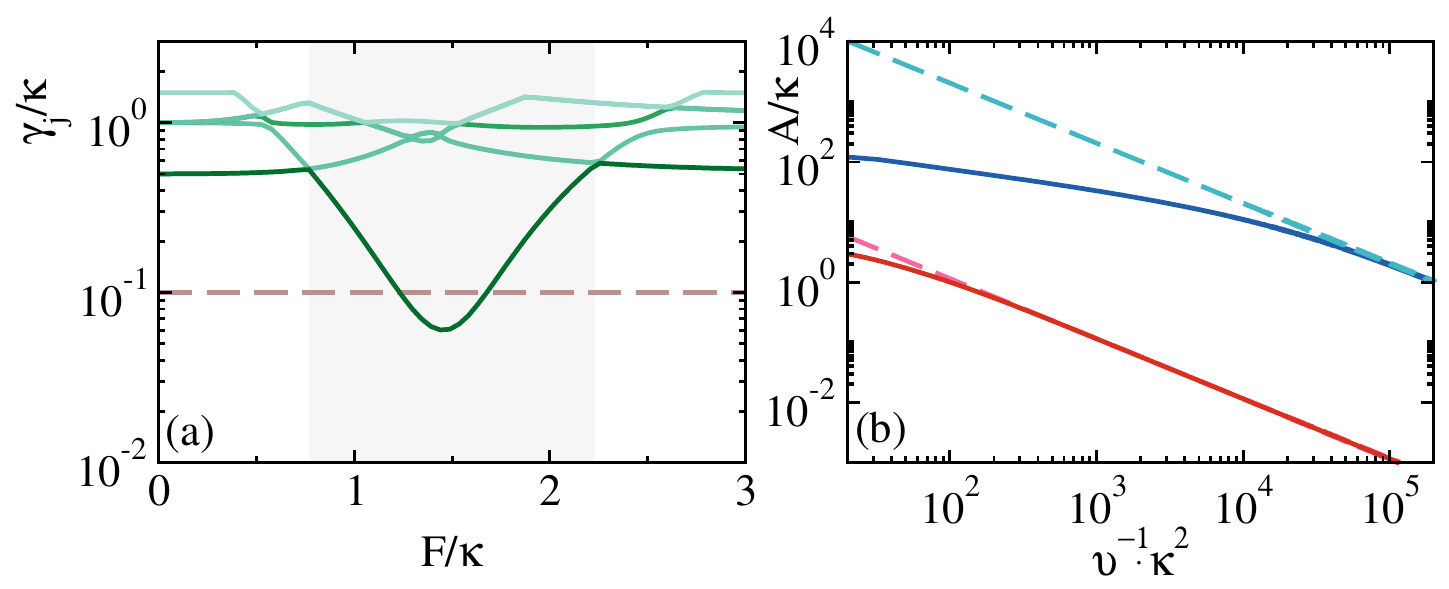}
\caption{(a) Decay rates $\gamma_j$ from a few lowest Liouvillian eigenvalues enumerated in the ascending order.  {\it Dashed line}: $1/t_c = \kappa/10$, giving the step duration $t_c$ in a staircase protocol similar to that sketched in the upper panels of Fig.~3 in the main article, with $F$ instead of $\delta$. The cavity decay rate, the strength of nonlinearity and the field amplitude are taken the same as in some of the figures in Ref.~[35] of the main article, in particular, $\kappa=1$, $U=-0.5$ and $\delta=-2$. {\it Gray area} indicates the critical region of the width $F^{\textup{max}} - F^{\textup{min}} \!\approx\! 1.46\, \kappa$. (b) Hysteresis area $\cal{A}$ ({\it solid lines}) as a function of $v^{-1}$ while sweeping $\delta$ ({\it blue}) and $F$ ({\it red}). Sweep velocities are defined as a ratio of the width of the critical region and $N t_c$. {\it Dashed lines}: Corresponding quasiadiabatic approximation.}
\label{fig_HYST_suppl}
\end{suppfigure}
\section*{Supplementary Note 4. Sweeping the pumping field amplitude}

As is mentioned in Ref.~[46] of the main article, sweeping the  pumping field amplitude $F$ at  fixed value of $\delta$ leads to the undesirable effect of non-vanishing entropy even for values of $F$ taken far to the right from the critical region. In Fig.~S\ref{fig_HYST_suppl} we also compare the results of calculations from the main article obtained under sweeping the detuning parameter $\delta$ with those obtained by sweeping the amplitude $F$ across the corresponding critical region. Two main conclusions can be drawn from the comparison of these two different sweeping protocols carried out at the same value of the Kerr nonlinearity strength $U$: (i) the critical region (in the same units of $\kappa$) is substantially narrower and the closure of the Liouvillian gap is much less pronounced when sweeping  $F$; (ii) as the result, the area of hysteresis is two orders of magnitude smaller and the crossover to quasiadiabatic regime occurs much earlier, at the value of  $v$ which is by more than two orders of magnitude larger than the one needed to enter the quasiadiabatic regime while sweeping $\delta$.

\section*{Supplementary Note 5. Experimental extraction of connection components}

Fixing the observable gauge defined above, we propose to measure the hysteresis branches in the cavity occupation number  at three different values of the sweeping velocity $v_i$ ($i=1,2,3$). For the forward $(f)$ and backward $(b)$ sweeping, they are theoretically given by $n_i^{(f)}(r) = n^{\textup{st}}(r) + \chi_{i}^{(f)} (r)$ and $n_i^{(b)}(r)=n^{\textup{st}}(r)+ \chi_{i}^{(b)} (r)$, respectively.  Supposing that the stationary value $n^{\textup{st}}(r)$ is unknown, we can readily extract the following information from the measured dependencies $n_i^{(b,f)}(r)$:
\begin{align}
\Delta \chi_i & \equiv \chi_{i}^{(b)} -\chi_{i}^{(f)} = n_i^{(b)} (r) - n_i^{(f)} (r), \\
\Delta \dot{\chi}_i &= \frac{d}{dr} [n_i^{(b)} (r) - n_i^{(f)} (r)].
\end{align}
Taking into account that each of $\chi_{i}^{(f,b)}(r)$ satisfies the equation (\ref{dchi1_dr}) with the corresponding $v_i$ and the sign choice, we straightforwardly derive the following relations
\begin{eqnarray}
v_i \Delta \dot{\chi}_i  = 2 \gamma_1 (r) \left[\bar{n}_i(r) - n^{\textup{st}}(r)\right] - A_{11}(r) v_i \Delta \chi_i ,
\label{gamA11}
\end{eqnarray}
where $\bar{n}_i (r) = \frac12 [n_i^{(b)}(r)  + n_i^{(f)}(r)]$ are the quantities already known from the measurements. It further follows
\begin{eqnarray}
v_1 \Delta \dot{\chi}_1 - v_2 \Delta \dot{\chi}_2= 2 \gamma_1(r) \left[\bar{n}_1(r)  - \bar{n}_2(r)  \right] - A_{11}(r) \left( v_1 \Delta \chi_1 - v_2 \Delta \chi_2 \right), \\
v_1 \Delta \dot{\chi}_1 - v_3 \Delta \dot{\chi}_3= 2 \gamma_1(r) \left[\bar{n}_1(r)  - \bar{n}_3(r)  \right] - A_{11}(r) \left( v_1 \Delta \chi_1 - v_3 \Delta \chi_3 \right).
\end{eqnarray}
Solving these equations we establish $\gamma_1(r)$ and $A_{11}(r)$. Plugging the obtained results into Eq.~(\ref{gamA11}) we, in turn, extract $n^{\textup{st}}(r)$, and hence $ \dot{n}^{\textup{st}}(r)$. Then, it is possible to establish $\chi_{i}^{(f)}(r)$ and $\dot{\chi}_{i}^{(f)}(r)$ [or $\chi_{i}^{(b)}(r)$ and $\dot{\chi}_{i}^{(b)}(r)$]. Inserting these dependencies into Eq.~(\ref{dchi1_dr}) we finally evaluate $A_{10}(r)$. 

Thus, from the three independent measurements one can find the relevant components of the Sarandy-Lidar connection.

\end{widetext}

\begin{thebibliography}{99}
\bibitem{berry84}
M. V. Berry, {\it Quantal phase factors accompanying adiabatic changes}, \href{https://doi.org/10.1098/rspa.1984.0023}{Proc. R. Soc. London A {\bf 392}, 45 (1984).}

\bibitem{berry89}
M. V. Berry, {\it The quantum phase, five years after}, in {\it Geometric phases in physics}, edited by A. Shapere and F. Wilczek, Advanced Series in Mathematical Physics Vol. 5 (World Scientific, Singapore, 1989).

\bibitem{simon83}
B. Simon, {\it Holonomy, the quantum adiabatic theorem, and Berry's phase}, \href{https://doi.org/10.1103/PhysRevLett.51.2167}{Phys. Rev. Lett. {\bf 51}, 2167 (1983).}

\bibitem{altzirn}
A. Altland and M. R. Zirnbauer, {\it Nonstandard symmetry classes in mesoscopic normal-superconducting hybrid structures}, \href{https://doi.org/10.1103/PhysRevB.55.1142}{Phys. Rev. B {\bf 55}, 1142 (1997).}

\bibitem{ryu}
S. Ryu, A. Schnyder, A. Furusaki, and A. Ludwig, {\it Topological insulators and superconductors: Tenfold way and dimensional hierarchy}, \href{https://doi.org/10.1088/1367-2630/12/6/065010}{New J. Phys. {\bf 12}, 065010 (2010).}

\bibitem{hasan2010}
M. Z. Hasan and C. L. Kane, {\it Colloquium: Topological insulators}, \href{https://doi.org/10.1103/RevModPhys.82.3045}{Rev. Mod. Phys. {\bf 82}, 3045 (2010).}

\bibitem{qi2011}
X.-L. Qi and S.-C. Zhang, {\it Topological insulators and superconductors}, \href{https://doi.org/10.1103/RevModPhys.83.1057}{Rev. Mod. Phys. {\bf 83}, 1057 (2011).}

\bibitem{schultz2012}
T. Schulz, R. Ritz, A. Bauer, M. Halder, M. Wagner, C. Franz, C. Pfleiderer, K. Everschor, M. Garst, and  A. Rosch, {\it Emergent electrodynamics of skyrmions in a chiral magnet}, \href{https://doi.org/10.1038/nphys2231}{Nat. Phys. {\bf 8}, 301 (2012).}
 
\bibitem{braun2012}
H. B. Braun, {\it Topological effects in nanomagnetism: From superparamagnetism to chiral quantum solitons}, \href{https://doi.org/10.1080/00018732.2012.663070}{Adv. Phys. {\bf 61}, 1 (2012).}

\bibitem{berg2014}
K. von Bergmann, A. Kubetzka, O. Pietzsch, and R. Wiesendanger, {\it Interface-induced chiral domain walls, spin spirals and skyrmions revealed by spin-polarized scanning tunneling microscopy}, \href{https://doi.org/10.1088/0953-8984/26/39/394002}{J. Phys. Condens. Matter {\bf 26}, 394002 (2014).}
 
\bibitem{wang2009}
Z. Wang, Y. Chong, J. D. Joannopoulos, and M. Soljacic, {\it Observation of unidirectional backscattering-immune topological electromagnetic states}, \href{https://doi.org/10.1038/nature08293}{Nature (London) {\bf 461}, 772 (2009).}

\bibitem{hafezi2013}
M. Hafezi, S. Mittal, J. Fan, A. Migdall, and J. M. Taylor, {\it Imaging topological edge states in silicon photonics}, \href{https://doi.org/10.1038/nphoton.2013.274}{Nat. Photonics {\bf 7}, 1001 (2013).}

\bibitem{Lujoann2014}
L. Lu, J. D. Joannopoulos, and M. Soljacic, {\it Topological photonics}, \href{https://doi.org/10.1038/nphoton.2014.248}{Nat. Photonics {\bf 8}, 821 (2014).}

\bibitem{eichen2009}
M. Eichenfield, J. Chan, R. M. Camacho, K. J. Vahala, and O. Painter, {\it Optomechanical crystals}, \href{https://doi.org/10.1038/nature08524}{Nature (London) {\bf 462}, 78 (2009).}
  
 \bibitem{safavi2014}
A. H. Safavi-Naeini, J. T. Hill, S. Meenehan, J. Chan, S. Gr{\"o}blacher, and O. Painter,
{\it  Two-dimensional phononic-photonic band gap optomechanical crystal cavity}, \href{https://doi.org/10.1103/PhysRevLett.112.153603}{Phys. Rev. Lett. {\bf 112}, 153603 (2014).}
  
\bibitem{ls1}
M. S. Sarandy and D. A. Lidar, {\it Adiabatic approximation in open quantum systems}, \href{https://doi.org/10.1103/PhysRevA.71.012331}{Phys. Rev. A {\bf 71}, 012331 (2005).}
  
\bibitem{ls2}
M. S. Sarandy and D. A. Lidar, {\it Abelian and non-Abelian geometric phases in adiabatic open quantum systems}, \href{https://doi.org/10.1103/PhysRevA.73.062101}{Phys. Rev. A {\bf 73}, 062101 (2006).}

\bibitem{weg1}
T. Pluecker, M. R. Wegewijs, and J. Splettstoesser, {\it Gauge freedom in observables and Landsberg's nonadiabatic geometric phase: Pumping spectroscopy of interacting open quantum systems}, \href{https://doi.org/10.1103/PhysRevB.95.155431}{Phys. Rev. B {\bf 95}, 155431 (2017).}

\bibitem{AvronJStatPhys}
J. E. Avron, M. Fraas, G. M. Graf,
{\it Adiabatic Response for Lindblad Dynamics}, \href{https://doi.org/10.1007/s10955-012-0550-6}{J. Stat. Phys. {\bf 148}, 800 (2012).}

\bibitem{albertPRX}
V. V. Albert, B. Bradlyn, M. Fraas, and L. Jiang, {\it Geometry and response of Lindbladians}, \href{https://doi.org/10.1103/PhysRevX.6.041031}{Phys. Rev. X {\bf 6}, 041031 (2016).}

\bibitem{weg3}
F. Reckermann, J. Splettstoesser, and M. R. Wegewijs, {\it Interaction-induced adiabatic nonlinear transport}, \href{https://doi.org/10.1103/PhysRevLett.104.226803}{Phys. Rev. Lett. {\bf 104}, 226803 (2010).}

\bibitem{diehl}
S. Diehl, E. Rico, M. A. Baranov, and P. Zoller, {\it Topology by dissipation in atomic quantum wires}, \href{https://doi.org/10.1038/nphys2106}{Nat. Phys. {\bf 7}, 971 (2011).}

\bibitem{Rabl2015}
T. J. Milburn, J. Doppler, C. A. Holmes, S. Portolan, S. Rotter, and P. Rabl, {\it General description of quasiadiabatic dynamical
phenomena near exceptional points}, \href{https://doi.org/10.1103/PhysRevA.92.052124}{Phys. Rev. A {\bf 92}, 052124 (2015).}

\bibitem{Rotter2016}
J. Doppler, A. A. Mailybaev, J. B\"{o}hm, U. Kuhl, A. Girschik, F. Libisch, T. J. Milburn, P. Rabl, N. Moiseyev, and S. Rotter, {\it Dynamically encircling an exceptional point for asymmetric mode switching}, \href{https://doi.org/doi:10.1038/nature18605}{Nature {\bf 537}, 76, (2016).}

\bibitem{carmichael}
H. J. Carmichael, {\it Breakdown of photon blockade: A dissipative quantum phase transition in zero dimensions}, \href{https://doi.org/10.1103/PhysRevX.5.031028}{Phys. Rev. X {\bf 5}, 031028 (2015).}
  
\bibitem{DW}
P. D. Drummond and D. F. Walls, {\it Quantum theory of optical bistability. I: Nonlinear
polarisability model}, \href{https://doi.org/10.1088/0305-4470/13/2/034}{J. Phys. A: Math. Gen. {\bf 13}, 725 (1980).}
  
\bibitem{plenio}
 A. Le Boit\'e, M.-J. Hwang, and M.B. Plenio, {\it Metastability in the driven-dissipative Rabi model}, \href{https://doi.org/10.1103/PhysRevA.95.023829}{Phys. Rev. A {\bf 95}, 023829 (2017).}
 
\bibitem{Macieszczak:2016aa}
K. Macieszczak, M. Guta, I. Lesanovsky, and J. P. Garrahan, {\it  Towards a Theory of Metastability in Open Quantum Dynamics}, \href{https://doi.org/10.1103/PhysRevLett.116.240404}{Phys. Rev. Lett. {\bf 116}, 240404 (2016).}

\bibitem{Minganti:2018aa}
F. Minganti, A. Biella, N. Bartolo, and C. Ciuti, {\it  Spectral theory of Liouvillians for dissipative phase transitions}, \href{https://doi.org/10.1103/PhysRevA.98.042118}{Phys. Rev. A {\bf 98}, 042118 (2018).}

\bibitem{Armen2006}
 M. A. Armen and H. Mabuchi,  {\it Low-lying bifurcations in cavity quantum electrodynamics}, \href{10.1103/PhysRevA.73.063801}{Phys. Rev. A {\bf 73}, 063801 (2006).}
 
\bibitem{siddiqi}
V. E. Manucharyan, E. Boaknin, M. Metcalfe, R. Vijay, I. Siddiqi, and M. Devoret,
{\it Microwave bifurcation of a Josephson junction: Embedding-circuit requirements},
\href{https://doi.org/10.1103/PhysRevB.76.014524}{Phys. Rev. B {\bf 76}, 014524 (2007).}
 
\bibitem{shumeiko}
W. Wustmann and V. Shumeiko, {\it Parametric resonance in tunable superconducting cavities}, \href{https://doi.org/10.1103/PhysRevB.87.023829}{Phys. Rev. B {\bf 87}, 184501 (2013).}
 
\bibitem{Angerer2017} 
A. Angerer,  S. Putz, D. O. Krimer, T. Astner, M. Zens, R. Glattauer, K. Streltsov, W. J. Munro, K. Nemoto, S. Rotter, J.  Schmiedmayer, and J. Majer, {\it Dynamical exploration of amplitude bistability in engineered quantum systems}, \href{https://doi.org/10.1126/sciadv.1701626}{Science Adv. {\bf 3}, e1701626 (2017).}
 
\bibitem{Krimer:2019aa} 
D. O. Krimer, M. Zens, and S. Rotter, Critical phenomena and nonlinear dynamics in a spin ensemble strongly coupled to a cavity.
I. Semiclassical approach, \href{https://doi.org/10.1103/PhysRevA.100.013855}{Phys. Rev. A {\bf 100}, 013855 (2019)}. 
 
\bibitem{ciuti1}
W. Casteels, F. Storme, A. Le Boit\'e, and C. Ciuti, {\it Power laws in the dynamic hysteresis of quantum nonlinear photonic resonators}, \href{https://doi.org/10.1103/PhysRevA.93.033824}{Phys. Rev. A {\bf 93}, 033824 (2016).}
 
\bibitem{Casteels:2017aa} 
W. Casteels, R. Fazio, and C. Ciuti,  {\it Critical dynamical properties of a first-order dissipative phase transition}, \href{https://doi.org/10.1103/PhysRevA.95.012128}{Phys. Rev. A {\bf 95}, 012128 (2017).}
 
\bibitem{devoret}
I. Siddiqi, R. Vijay, F. Pierre, C. M. Wilson, M. Metcalfe, C. Rigetti, L. Frunzio, and M. H. Devoret, {\it rf-Driven Josephson bifurcation amplifier for quantum measurement}, \href{https://doi.org/10.1103/PhysRevLett.93.207002}{Phys. Rev. Lett. 93, 207002 (2004).}
  
\bibitem{reimer}
V. Reimer, K. Pedersen, N. Tanger, M. Pletyukhov, and V. Gritsev, {\it Nonadiabatic effects in periodically driven dissipative open quantum systems}, \href{https://doi.org/10.1103/PhysRevA.97.043851}{Phys. Rev. A {\bf 97}, 043851 (2018).}
  
\bibitem{ciuti2}
S. R. K. Rodriguez, W. Casteels, F. Storme, N. Carlon Zambon, I. Sagnes, L. Le Gratiet, E. Galopin, A. Lemaitre, A. Amo, C. Ciuti, and J. Bloch, {\it Probing a dissipative phase transition via dynamical optical hysteresis}, \href{https://doi.org/10.1103/PhysRevLett.118.247402}{Phys. Rev. Lett. {\bf 118}, 247402 (2017).}

\bibitem{foot2}
We call the regime noncritical, whenever $1/t_c \ll  \gamma_{q\geq 1}$.

\bibitem{foot3}
W.l.o.g. we assume that $| \rho (t=0)) = |\rho^0_{\delta_{\textup{min}}})$.

\bibitem{foot4}
A truncation of the higher Liouvillian modes with $q \geq 2$ relying  on the condition $e^{- t_c \gamma_{q}} \ll 1$  provides a description of metastable states in terms of just two Liouvillian modes $| \rho^0)$ and $| \rho^1 )$. Additionally, we note that $\omega_1$ identically vanishes in the critical regime \cite{ciuti1}.


\bibitem{suppl}
See Supplemental Material below, where we briefly sketch the derivation of the master equation and the eigenvalue problem, present details on a simplified description of metastable states and explain how the connection components can be experimentally extracted. The results of calculations on sweeping the pumping field amplitude are also discussed.

\bibitem{lands}
A. S. Landsberg, {\it Geometrical phases and symmetries in dissipative systems}, \href{https://doi.org/10.1103/PhysRevLett.69.865}{Phys. Rev. Lett. {\bf 69}, 865 (1992).}

\bibitem{ning}
C. Z. Ning and H. Haken, {\it Geometrical phase and amplitude accumulations in dissipative systems with cyclic attractors}, \href{https://doi.org/10.1103/PhysRevLett.68.2109}{Phys. Rev. Lett. {\bf 68}, 2109 (1992).}

\bibitem{foot5}
It is worth noting that sweeping the drive amplitude $F$ at a fixed value of $\delta$ not only leads to a much weaker closure of the Liouvillian gap but also  leads to the undesirable effect of non-vanishing entropy outside the critical region
as compared to the case when the detuning parameter $\delta$ is swept. As a consequence, geometrical effects related to the hysteretic area become much less pronounced (see also  \cite{suppl} for more details).

\bibitem{Xiang2013}
Z.-L. Xiang, S. Ashhab, J. Q. You, and F. Nori,  {\it Hybrid quantum circuits: Superconducting circuits interacting with other quantum systems}, \href{https://doi.org/10.1103/RevModPhys.85.623}{Rev. Mod. Phys. {\bf 85}, 623 (2013)}.

\bibitem{Kurizki2015}
G. Kurizki, P. Bertet, Y. Kubo, K. M\o lmer, D. Petrosyan, P. Rabl, and J. Schmiedmayer, {\it Quantum technologies with hybrid systems}, \href{https://doi.org/10.1073/pnas.1419326112}{Proceedings of the National Academy of Sciences {\bf 112}, 3866 (2015).}

\bibitem{Letscher:2017aa}
F. Letscher, O. Thomas, T. Niederpr\"um, M. Fleischhauer, and H. Ott, {\it Bistability versus metastability in driven dissipative Rydberg gases}, \href{https://doi.org/10.1103/PhysRevX.7.021020}{Phys. Rev. X {\bf 7}, 021020 (2017).}

\end{thebibliography}
\end{document}